\documentclass[prb,twocolumn,amsmath,showpacs]{revtex4-1}

\usepackage{graphicx}
\usepackage{braket}
\usepackage{mathrsfs}
\usepackage{pict2e}

\usepackage[unicode=true,colorlinks=true]{hyperref}
\usepackage{bm}
\newcommand{\bvec}[1]{{\mathbf{#1}}}
\newcommand{\bgreek}[1]{{\bm{{#1}}}}

\newcommand{\scatt}[1]{%
   \multiput(#1,0.0)(0.0,0.2){5}{\line(0,1){0.1}} %
   \put(#1,0.0){\Line(-0.1,-0.1)(0.1,0.1)} \put(#1,0.0){\Line(-0.1,0.1)(0.1,-0.1)} %
   \put(#1,1.0){\Line(-0.1,-0.1)(0.1,0.1)} \put(#1,1.0){\Line(-0.1,0.1)(0.1,-0.1)} %
}
\newcommand{\dline}[1]{%
   \put(#1,0.0){\Line(0.0,0.0)(1.0,0.0)} \put(#1,1.0){\Line(0.0,0.0)(1.0,0.0)} %
   \put(#1,0.0){\vector(1,0){0.6}} \put(#1,1.0){\vector(1,0){0.6}} %
}
\begin{document}

\title{Non-diffusion theory of weak localization in graphene}
\author{M.O. Nestoklon, N.S. Averkiev}
\affiliation{Ioffe Physical-Technical Institute, Russian Academy of Sciences, St. Petersburg 194021, Russia}
\begin{abstract}
    We put forward a theory of the weak localization in two dimensional 
    graphene layers which explains experimentally observable transition 
    between positive and negative magnetoresistance. 
    Calculations are performed for the whole range of classically weak magnetic field
    with account on intervalley transitions.
    Contribution to the quantum correction which stems from closed trajectories 
    with few scatterers is carefully taken into account.
    We show that intervalley transitions lead not only to the transition
    from weak antilocalization to the weak localization, but also to the 
    non-monotonous dependence of the conductivity on the magnetic field.
\end{abstract}


\maketitle

\section{Introduction}
Recent studies of perfect graphene layers gave a new impulse\cite{Novoselov,Neto} 
for experimental and theoretical 
investigation of two- and three-dimensional structures with linear energy spectrum of the carriers. 
Despite the differences in chemical bonding, electrical properties of graphene layers,
surface states in Bi$_2$Se$_3$ Bi$_2$Te$_3$, Bi$_2$Te$_2$Se
and quantum wells based on HgTe are 
defined mostly by the linear energy dependence on lateral wave vector.
This dependence leads to weak antilocalization (WAL) and positive magnetoresistance 
in classically weak magnetic fields if all relaxation processes take place inside 
one dispersion cone.\cite{Tkachov13,KachorovskiiPSS,Kachorovskii14} 
As long as zero energy in graphene is located at the Brillouin zone boundary in $K$ point
in contrast to topological insulators, significant role in transport phenomena is also played by intervalley 
transitions between equivalent points $K$ and $K'$. When considering the intervalley transitions, 
it is important to take into consideration that valleys in graphene are connected by time inversion.
It leads\cite{Ando02,McCann,OurEPL} to weak localization (WL) when the phase
relaxation time $\tau_{\phi}$ in each valley is larger than intervalley transition time $\tau_v$.
The ratio $\tau_{\phi}/\tau_v$
in graphene may be controlled by changing the gate bias. 
Previous theoretical investigations considered either non-diffusion theory with
account on intervalley transitions in zero magnetic field \cite{OurEPL} or magnetoresistance in 
diffusion regime.\cite{McCann,Kechedzhi07} Experimentally, magnetoresistance has been extencively 
studied \cite{Tikhonenko09,Nath12,Baker12}
and it has been shown that both (WAL and WL) regimes are possible depending on technology of sample 
and applied bias.

Weak localization phenomenon is based on the change of carrier return probability due to
interference of the waves travelling the same path in the opposite directions. Magnetic field
applied to the structure
or other processes of the phase decoherence change the interference conditions 
which results in changing 
the contribution to the conductivity from the closed trajectories. As long as the magnetic 
field which changes the phase is classically weak and the change of wavefunction phase is quantum 
phenomenon, such corrections are normally called quantum corrections.

Intervalley transitions were considered in the framework of diffusion
approximation for graphene\cite{McCann} and for tellurium which has similar
bandstructure.\cite{Averkiev96} 
These results show that transition from antilocalization to localization 
regime in graphene is possible when 
$3\ln(\tau_v/\tau_{\phi})>2\ln(\tau_{\phi}/\tau_{tr})$ 
(here $\tau_{tr}$ is momentum relaxation time). 
Both terms should be significantly large than unity.
The diffusion theory describes the quantum correction to the conductivity in the limit 
$\ln(\tau_{\phi}/\tau_{tr})\gg 1$ assuming that this it is relatively easy to go beyond 
this regime in highly conducting samples.
To overcome this limitation in theory 
it is necessary to include into consideration closed paths with small number of
scatterers. Movement along such trajectories is non-diffusion.
As long as in reality both $\ln(\tau_v/\tau_{\phi})$ and $\ln(\tau_{\phi}/\tau_{tr})$
are less than $10$, diffusion theory may give only qualitative estimation
of the quantum correction and one may not use its results to analyze the 
magnetoresistance caused by the weak localization.

The weak localization regime is known to be protected by the 
time inversion.\cite{Kechedzhi07}
In the system where carriers are located near $\Gamma$ point of Brillouin Zone
the time inversion 
guarantees the diffusion pole for the Cooperons because the correlator in 
self-energy part coincides with the correlator in Cooperon equation. 
In multivaley systems where the valleys are connected by time inversion
(see e.g. Ref.~\onlinecite{Tarasenko07}) time inversion guarantees diffusion 
pole for intervalley Cooperons only. For diffusion pole for intravalley
Cooperons additional symmetries (in graphene space inversion inside one valley
\cite{McCann,Kechedzhi07}) are needed.

Generalization of the theory of magnetic field quantum corrections to the 
non-diffusion case appears to be the last conceptual theoretical problem
of the weak localization theory in graphene. 
This theory may also be considered as a limit case of strong spin-orbit
interaction (of Rashba or Dresselhaus type) in two-dimensional 
electron systems,\cite{Glazov09} when linear in ${\bf k}$ terms in Hamiltonian dominate.
The goal of this work is the theoretical 
study of the weak localization in graphene in the full range of classically weak magnetic fields.

The manucript is organized as follows:
Section \ref{SEC:QAWL} gives an extended introduction in the weak localization in graphene.
In section \ref{SEC:theory} we give some details of the non-diffusion calculations of the weak 
localization: we start from the main starting points of the theory, choice of basis functions, 
Hamiltonian, scattering matrix elements, etc. In subsection \ref{SSEC:G} we derive the Green 
function in two valleys in real space, in subsection \ref{SSEC:C} we write and solve Cooperon 
equation, subsection \ref{SSEC:sigma} gives the derivation of equations for weak localization 
corrections. Subsection \ref{SSEC:LowH} gives low magnetic field limit of the results obtained
before. Finally, in section \ref{SEC:res} we present the results of the weak 
localization correction computations. 
In addition, in appendix \ref{SEC:P} we give important details of numerical calculation of integrals 
which arise in computation of weak localizations correction, appendix \ref{SEC:sum} gives 
a recipe to simplify calculation of infinite sums for the weak localization correction and 
finally appendix \ref{SEC:int} gives some mathematical relations used in the 
manuscript.

\section{Qualitative analysis of the weak (anti)localization}\label{SEC:QAWL}
The intra-valley electron scattering from a symmetric
short-range (as compared to de Broglie wavelength) impurity 
in graphene is described by the matrix element of scattering
\begin{equation}\label{eq:scatter_intra}
V( {\bf k}', {\bf k}) \propto {\rm e}^{i (\varphi-\varphi')/2} \cos [(\varphi-\varphi')/2] \:,
\end{equation}
where ${\bf k}=(k\cos\varphi,k\sin\varphi)$ and ${\bf k}'=(k\cos\varphi',k\sin\varphi')$ 
are wave vectors of respectively incident and scattered electrons.

One can see that the direct back scattering from an impurity is suppressed and, what is 
more important for quantum effects, the scattering introduces the phase 
$(\varphi-\varphi')/2$ to the electron wave function. Therefore, an electron traveling 
clockwise along a closed path and finally scattered back gains the additional phase 
$\pi/2$ while the electron traveling in the opposite direction gains the phase 
$-\pi/2$. The phase shift of $\pi$ between these two waves results in a destructive 
interference and, hence, in the antilocalization of carriers.

Other forms of scattering amplitude lead to 
the phase gain which depends on the particular trajectory even for closed paths. 
Averaging over the trajectories destroys the wave interference and results in no quantum
corrections to conductivity (see Refs.~\onlinecite{Morozov,McCann,Morpurgo}).
It is a general rule which manifests itself in a fact that 
corrections to electron Hamiltonian due to e.g. trigonal warping, 
nonsymmetric scattering, etc.
in graphene suppress the weak antilocalization.\cite{McCann} 

In the system where carriers are located near $\Gamma$ point of Brillouin Zone
the time inversion 
guarantees the diffusion pole for the intravalley Cooperons as long as correlator in 
self-energy part coincides with the correlator in Cooperon equation. 
In multivaley systems where the valleys are connected by time inversion
(see e.g. Ref.~\onlinecite{Tarasenko07}) time inversion guarantees diffusion 
pole only for intervalley Cooperons which do not contribute to conductivity 
in the absence of intervalley transitions.
For diffusion pole for intravalley Cooperons other symmetries 
(in graphene space inversion inside one valley \cite{McCann,Kechedzhi07}) are needed.
In the presence of intervalley scattering, Cooperons associated 
with such scattering contribute to conductivity giving 
rise to weak localization.\cite{McCann,OurEPL}

Intervalley contribution to the conductivity due to its time-invariant nature
results in a conventional WL, as in spinless single valley case. 
However, it is proportional to the intervalley scattering rate.
Changing intervalley scattering rate one may continously switch between
two cases.\cite{Ando02,McCann,OurEPL}

\section{Theory}\label{SEC:theory}
In the following we work in the basis $\{KA,KB,K'B,K'A\}$ with basis functions 
$KA,KB$ in one valley transform as $x\pm i y$  and the basis functions in the 
second valley are obtained by $C_2$ rotation perpendicular to graphene 
sheet.\cite{Winkler10} In this basis Hamiltonian is 
\begin{equation}\label{eq:H}
  \mathscr{H} = \hbar v \begin{pmatrix}
    \bgreek{\sigma}\cdot\bvec{k} &  0 \\ 0 & -\bgreek{\sigma}\cdot\bvec{k}
  \end{pmatrix}\;.
\end{equation}
In the magnetic field, we neglect zeeman-like terms which do not contribute to the weak localization 
and the Hamiltonian reads 
\begin{equation}
 \mathscr{H} = \hbar v \frac{\sqrt{2}}{\ell_B}
 \begin{pmatrix} 
 0 & a_- & 0 & 0 \\
 a_+ & 0 & 0 & 0 \\
 0 & 0 & 0 & -a_- \\
 0 & 0 & -a_+ & 0 
 \end{pmatrix}\;,
\end{equation}
where we defined standard ladder operators $a_{\pm} = \ell_B(k_x\pm i k_y)/\sqrt{2}$,
and $\ell_B$ is magnetic length.
This Hamiltonian gives us the positive energy solutions in two valleys:
\begin{equation}\label{eq:functions}
\begin{split}
  \Psi_{N,k,1} (\bvec{r}) &= \frac1{\sqrt2} \begin{pmatrix} 
    \psi_{N-1,k}(\bvec{r}) \\ \psi_{N,k}(\bvec{r}) \\ 0 \\ 0
  \end{pmatrix}\;,
  \\
  \Psi_{N,k,2} (\bvec{r}) &= \frac1{\sqrt2} \begin{pmatrix} 
    0 \\ 0 \\ \psi_{N-1,k}(\bvec{r}) \\ -\psi_{N,k}(\bvec{r})
  \end{pmatrix}\;.
\end{split}
\end{equation}
where $\psi_{N,k}$ are functions of electron in magnetic field in Landau gauge
\begin{equation}\label{eq:osc_fun}
\begin{split}
    \psi_{N,k}(\bvec{r}) &= 
    \frac{e^{iky-\frac{(x+\ell_B^2k)^2}{2\ell_B^2}}}{\sqrt{\ell_B}\sqrt{2^NN!\sqrt{\pi}}}
  H_N\left(\frac{x+\ell_B^2k}{\ell_B}\right),\\
  H_N(\xi) &= (-1)^Ne^{\xi^2}\frac{d^N}{d\xi^N}e^{-\xi^2}
\end{split}
\end{equation}
\begin{equation}
\begin{split}
  a_+ \psi_{N-1,k}(\bvec{r}) &= \sqrt{N} \psi_{N,k}(\bvec{r}),\\
  a_- \psi_{N,k}(\bvec{r}) &= \sqrt{N} \psi_{N-1,k}(\bvec{r})
\end{split}
\end{equation}

Scattering may be obtained from symmetry considerations. 
Assuming the non-magnetic potential with symmetry $\Gamma_1^+$ in 
Koster notation \cite{Koster} with the center at $\bvec{r}_0$ it reads as 
\begin{subequations}\label{eq:scattering}
\begin{equation}
    \delta\mathcal{H}_{intra} (\bvec{r};\bvec{r}_0) =
    \sqrt{\frac{2v}{nk_F\tau}}
    \begin{pmatrix}  1 & 0 & 0 & 0 \\ 0 & 1 & 0 & 0 \\ 0 & 0 & 1 & 0 \\ 0 & 0 & 0 & 1
    \end{pmatrix}
    \delta(\bvec{r}-\bvec{r}_0),
\end{equation}
\begin{equation}
    \delta\mathcal{H}_{inter} (\bvec{r};\bvec{r}_0) =
    \sqrt{\frac{2v}{nk_F\tau_v}}
    \begin{pmatrix}  0 & 0 & \varepsilon & 0 \\ 0 & 0 & 0 & \varepsilon \\ 
    \varepsilon^* & 0 & 0 & 0 \\ 0 & \varepsilon^* & 0 & 0 
    \end{pmatrix} 
    \delta(\bvec{r}-\bvec{r}_0),
\end{equation}
\end{subequations}
where $\tau$ is quantum relaxation time and 
$\varepsilon=e^{i(\bvec{K}-\bvec{K}')\cdot\bvec{r}_0}$ is a phase which stems from difference
of valley positions in $k$-space.

Note that in Ref.~\onlinecite{OurEPL} authors omitted phase factor which takes into 
account position of impurity.
\subsection{Green function}\label{SSEC:G}
Without account on scattering, the Green function reads
\begin{equation}\label{EQ:G0def}
  G^{R,A}_0(\bvec{r},\bvec{r}') = \sum_{N,k,t} 
  \frac{\Psi_{N,k,t}(\bvec{r}) \Psi_{N,k,t}^+(\bvec{r}')}
  {E_F-\varepsilon_N\pm i \frac{\hbar}{2\tau_{\phi}}}
\end{equation}
Here $t$ is valley index. 

Solution of the Dyson equation for the renormalized Green function 
for the given short range scatterers \eqref{eq:scattering} in the
first order of scatterers density reads as
\begin{equation}\label{EQ:Gdef}
  G^{R,A}(\bvec{r},\bvec{r}') = \sum_{N,k,t} 
  \frac{\Psi_{N,k,t}(\bvec{r}) \Psi_{N,k,t}^+(\bvec{r}')}
  {E_F-\varepsilon_N \pm i \frac{\hbar}{2\tau'}}
\end{equation}
where effective relaxation time $1/\tau' = 1/\tau_{\phi}+1/\tau + 1/\tau_v$ is 
defined by harmonic sum of phase relaxation time $\tau_{\phi}$, quantum relaxation 
time $\tau$ and intervalley transition time $\tau_v$.

With the help of results presented in Appendix~\ref{SEC:int} assuming $k_F\ell\gg1$ 
(where $k_F=E_F/\hbar v$ is wave vector at Fermi level)
in classically weak magnetic fields $k_F\ell_B\gg1$ at sufficiently small distances 
the effect of the magnetic field on a Green function may be written as a 
phase factor: 
\begin{equation}
  G^{R,A}({\bf{r}},{\bf{r}}') = 
  \exp{\left[ -i\frac{(x+x')(y-y')}{2\ell_B^2} \right]} G^{R,A}_{B0}( {\bf{r}}-{\bf{r}}' ) \:,
\end{equation}
where $G^{R,A}_{B0}( {\bf{r}}-{\bf{r}}' )$ are the Green functions of an electron in graphene
at zero field,
\begin{equation}\label{eq:GB0}
G^{R,A}_{B0}({\boldsymbol{\rho}}) = 
- \frac{ \exp[-\rho/(2\ell') \pm i ( k_F \rho + \pi/4) ]}{\sqrt{ 2\pi \rho / k_F} \: \hbar v}  \, g^{R,A}({\boldsymbol{\rho}}) \:,
\end{equation}
\[
g^{R,A}({\boldsymbol{\rho}}) =  \frac12
\begin{pmatrix}
  1        & \pm in_- & 0        & 0 \\
  \mp in_+ & 1        & 0        & 0 \\
  0        & 0        & 1        & \mp in_- \\
  0        & 0        & \pm in_+ & 1 
\end{pmatrix} \:,
 \]
$\ell'=\ell/\left(1+\tau/\tau_{\phi}+\tau/\tau_v\right)$, $\ell=v \tau$ is the mean free path.

\subsection{Cooperon equation}\label{SSEC:C}

The key point in calculation of the weak localization correction is the solution of Cooperon
equation which describes the sum of the fan diagrams (see Fig.~\ref{fig:Ceq}).
As we have a complicated structure of the Green function, to simplify the solution we rewrite 
Cooperon equation in the matrix form.

\begin{figure}
   \begin{picture}(6.5,1.5)(0.0,-0.25)
   \thicklines
   \put(0.0,0.35){$C^{(2)}=$}
   \scatt{1.05} \dline{1.05} \scatt{2.05}
   \put(2.40,0.35){$+$}
   \scatt{3.05} \dline{3.05} \scatt{4.05} \dline{4.05} \scatt{5.05}
   \put(5.40,0.35){$+ ... =$}
   \end{picture}
   \[
   = SPS + SPSPS +  ... = SP\cdot\left( \sum_{n=0}^{\infty} S(PS)^n \right) = 
   SP \cdot C
   \]
\caption{Illustration of the Cooperon equation} \label{fig:Ceq}
\end{figure}

We start from writing Cooperon equation in real space: 
\begin{equation}\label{eq:Cr16}
   C^{\alpha \beta}_{\gamma \delta} (\bvec{r},\bvec{r}') = 
   S^{\alpha \beta}_{\gamma \delta}(\bvec{r}) \delta(\bvec{r}-\bvec{r}') +
  \int 
     \left[SP\right]^{\alpha \zeta}_{\gamma \xi}(\bvec{r},\bvec{r}'') 
   C^{\zeta\beta}_{\xi\delta}(\bvec{r}'',\bvec{r}') d\bvec{r}'',
\end{equation}
where
\begin{equation}\label{eq:P_def}
   P^{\alpha \beta}_{\gamma \delta}(\bvec{r},\bvec{r}') = 
   G^A_{\alpha\beta}(\bvec{r},\bvec{r}')
   G^R_{\gamma \delta}(\bvec{r},\bvec{r}')
\end{equation}
and scattering correlator
\begin{equation}
   \left\langle
   \delta\mathcal{H}_{\alpha\zeta}(\bvec{r}-\bvec{r}_1)
   \delta\mathcal{H}_{\gamma\xi}(\bvec{r}'-\bvec{r}_2)
   \right\rangle_{\bvec{r}_1,\bvec{r}_2}
   = 
   S^{\alpha \zeta}_{\gamma \xi} \delta(\bvec{r}-\bvec{r'}).
\end{equation}

It is not convenient to solve Cooperon equation in this form. To solve it 
efficiently we have to transform it into matrix equation.

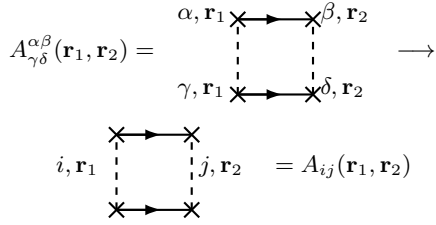
\begin{figure}
    $A^{\alpha\beta}_{\gamma\delta}(\bvec{r}_1,\bvec{r}_2)=$
    \raisebox{-0.5cm}{%
    \begin{picture}(3,1.5)(0.0,0.0)
    \thicklines
    \scatt{1.0} \dline{1.0} \scatt{2.0}
    \put(0.2,1.0){$\alpha,\bvec{r}_1$}
    \put(2.1,1.0){\rlap{$\beta,\bvec{r}_2$}}
    \put(0.2,0.0){$\gamma,\bvec{r}_1$}
    \put(2.1,0.0){\rlap{$\delta,\bvec{r}_2$}}
    \end{picture}} 
    $\longrightarrow$
    \raisebox{-0.5cm}{
    \begin{picture}(3,1.5)(0.0,0.0)
    \thicklines
    \scatt{1.0} \dline{1.0} \scatt{2.0}
    \put(0.2,0.5){$i,\bvec{r}_1$}
    \put(2.1,0.5){\rlap{$j,\bvec{r}_2$}}
    \end{picture}}
    $=A_{ij}(\bvec{r}_1,\bvec{r}_2)$
\caption{Illustration of replacement of four-tail diagram with a block with two indexes 
in product space. Here $ {\alpha \atop \gamma} \to i$,
${\beta \atop \delta} \to j$} \label{fig:Fourtail}
\end{figure}

To transform summation over two indices into standard matrix multiplication one needs to assume 
definite basis in direct product space. Below we use convention given in 
Table~\ref{tbl:prod_basis}. To distinguish between equations written in original and direct 
product spaces, later we use Greek letter indices for original space of graphene 
Hamiltonmian and Latin letter indices for direct product space.
For the Cooperon equation (see later) it is more convenient to 
rewrite scattering in product space. 
As a general rule, we rewrite each four-tail diagram with four indexes associated with four tails
with a block having two indices each associated with a pair of tails:
one to the left and one to the right, see Fig.~\ref{fig:Fourtail}.

\begin{table}
\caption{Basis of direct product used in paper. Minus in the table is a shorthand for 
$\ket{i}=-\ket{\alpha}\ket{\beta}$.}
\begin{tabular}{c|cccc|cccc|cccc|cccc}
$\alpha$ &1&1&2&2&1&   1&2&   2&3& 3&    4&    4& 3&    3&    4& 4 \\ 
$\gamma$ &1&2&1&2&3&   4&3&   4&1& 2&    1&    2& 3&    4&    3& 4 \\\hline
$i     $ &1&2&3&4&5&$-$6&7&$-$8&9&10&$-$11&$-$12&13&$-$14&$-$15&16\\ 
\end{tabular}
\label{tbl:prod_basis}
\end{table}

Following this rule, we associate a correlator for the pair of scatterings 
with the matrix 
\begin{equation}
   S_{ij} = W\cdot
   \begin{pmatrix}
   1 & 0 & 0 & 0 \\
   0 & 1 & -\Pi_v \frac{\tau}{\tau_v} & 0 \\
   0 & -\Pi_v \frac{\tau}{\tau_v} & 1 & 0 \\
   0 & 0 & 0 & 1
   \end{pmatrix},
\end{equation}
\begin{equation}
  W = \frac{\hbar^2v}{k_F\tau},
\end{equation}
and the matrix $\Pi_v$ defined as 
$ \Pi_v = \mathrm{diag} \left\{ -1 , 1, 1, -1 \right\}$ .

Product of two Green functions \eqref{eq:P_def} in the product space basis
given in Table~\ref{tbl:prod_basis} in given by block-diagonal matrix with
$4\times 4$ blocks $P_{sv}(\bvec{r},\bvec{r}')$ similar 
to the single-valley case:\cite{OurSSC}
\begin{equation}\label{Cooperon_kernel}
P_{sv}(\bvec{r},\bvec{r}') = 
\frac{P_0({\bf{r}},{\bf{r}}')}2
\begin{bmatrix}
   1     & in_-  & -in_-   &   n_-^2 \\
 -in_+   &  1    &  -1     & -in_-   \\
  in_+   & -1    &   1     &  in_-   \\
   n_+^2 & in_+  & -in_+   &   1  
\end{bmatrix}\:,
\end{equation}
\begin{equation}\label{eq:P0}
P_0({\bf{r}},{\bf{r}}')=\frac{{\rm e}^{-|{\bf r}-{\bf r}'|/\ell'}}{2\pi \ell |{\bf r}-{\bf r}'|} \exp{\left[ -i\frac{(x+x')(y-y')}{\ell_B^2} \right]} \:.
\end{equation}
Due to the block structure of Cooperon equation kernel, we may rewrite a
single $16\times16$ matrix equation \eqref{eq:Cr16} as a system of equations for $4\times4$ blocks defined as
\begin{equation}\label{eq:Cr}
   C (\bvec{r},\bvec{r}') = 
   \begin{pmatrix}
   C_{0}(\bvec{r},\bvec{r}') & 0 & 0 & 0 \\
   0 & C_{1}(\bvec{r},\bvec{r}') &-C_{2}(\bvec{r},\bvec{r}') & 0 \\
   0 &-C_{2}(\bvec{r},\bvec{r}') & C_{1}(\bvec{r},\bvec{r}') & 0 \\
   0 & 0 & 0 & C_{0}(\bvec{r},\bvec{r}')
   \end{pmatrix}
\end{equation}
\begin{widetext}
The Cooperon equation separates into equations for intravalley contribution
\begin{subequations}\label{eq:Cr012}
\begin{equation}\label{eq:Cr0}
   C_0 (\bvec{r},\bvec{r}') =  W \delta(\bvec{r}-\bvec{r}') +
  \int 
     P_{sv} (\bvec{r},\bvec{r}') C_0(\bvec{r}'',\bvec{r}') d\bvec{r}'',
\end{equation}
and two intervalley equations
\begin{equation}\label{eq:Cr1}
   C_1 (\bvec{r},\bvec{r}') = W \delta(\bvec{r}-\bvec{r}') +
  \int 
     P_{sv} (\bvec{r},\bvec{r}') C_1(\bvec{r}'',\bvec{r}') d\bvec{r}''
+ \frac{\tau}{\tau_v} \Pi_v \int P_{sv} (\bvec{r},\bvec{r}') C_2(\bvec{r}'',\bvec{r}') d\bvec{r}''
\end{equation}
\begin{equation}\label{eq:Cr2}
   C_2 (\bvec{r},\bvec{r}') =  \frac{\tau}{\tau_v} \Pi_v W \delta(\bvec{r}-\bvec{r}') +
  \int 
     P_{sv} (\bvec{r},\bvec{r}') C_2(\bvec{r}'',\bvec{r}') d\bvec{r}''
+ \frac{\tau}{\tau_v} \Pi_v \int P_{sv} (\bvec{r},\bvec{r}') C_1(\bvec{r}'',\bvec{r}') d\bvec{r}''
\end{equation}
\end{subequations}
\end{widetext}

To solve integral equations \eqref{eq:Cr012} we use approach\cite{OurSSC} based on 
Kawabata theory\cite{Kawabata} and rewrite them in the basis
\begin{subequations}\label{eq:basisN}
\begin{align}
\Phi_{N \geq 1, k}({\bf{r}})=&
\frac1{\sqrt2}
\left( \begin{array}{cccc}
   0        & \sqrt2\phi_{N-1,k} &  0        & 0 \\
   \phi_{N,k} & 0       &  \phi_{N,k} & 0 \\
   \phi_{N,k} & 0       & -\phi_{N,k} & 0 \\
   0        &  0        &  0 &  -\sqrt2\phi_{N+1,k}
\end{array} \right) ,
\\
\Phi_{0, k}({\bf{r}})=&\frac1{\sqrt2}
\left( \begin{array}{cccc}
   0        &  0        & 0  & 0 \\
   \phi_{0,k} & 0      &  \phi_{0,k} & 0 \\
   \phi_{0,k} & 0      & -\phi_{0,k} & 0 \\
   0        &  \sqrt2\phi_{0,k} &         & -\sqrt2\phi_{1,k}
\end{array} \right) .
\end{align}
\end{subequations}
where $\phi_{N,k}$ are the oscillator functions of particle with double charge in 
the Landau gauge.

Product of two Green functions \eqref{eq:P_def} is convenient to write in the basis 
\eqref{eq:basisN} as
\begin{equation}\label{eq:P_in_basis}
P( {\bf{r}},{\bf{r}}' ) = \frac{\tau'}{\tau}
\sum_{N,k} \Phi_{Nk}({\bf{r}}) P_N \Phi_{Nk}^{\dag}({\bf{r}}') \:.
\end{equation}
Straightforward calculation with the use of results presented in Appendix~\ref{SEC:int}
gives the following result for the decomposition of \eqref{eq:P_def} in the basis \eqref{eq:basisN}:
\begin{subequations}\label{eq:PN}
\begin{align}
P_{N \geq 1} = &  \epsilon \begin{pmatrix}
0 & 0 & 0 & 0 \\
0 &  P_{N-1}^{0}  &  -i P_{N}^{1}   & -P_{N+1}^{2}  \\
0 & -i P_{N}^{1}  &  P_{N}^{0}  & -iP_{N+1}^{1} \\
0 & -P_{N+1}^{2}  &-iP_{N+1}^{1}&  P_{N+1}^{0}
\end{pmatrix} \epsilon,
\\
P_{0} = &  \epsilon \begin{pmatrix}
0 & 0 & 0 & 0 \\
0 & P_{0}^{0} & 0 & 0 \\
0 & 0 &  P_{0}^{0}  &-iP_{1}^{1} \\
0 & 0 &-iP_{1}^{1}&  P_{1}^{0}
\end{pmatrix} \epsilon.
\end{align}
\end{subequations}
Here for brievity we introduced auxiliary matrix
$ \epsilon = \mathrm{diag} \left\{ 1, 1/\sqrt2, 1, 1/\sqrt2 \right\}$.
Integrals $P_N^{M}$ are given by
\begin{multline}\label{eq:PNM}
P_N^{M} = \frac{\ell_B}{\ell'} \sqrt{\frac{(N-M)!}{N!}} \times \\
\int_0^{\infty} \exp{\left[ - x \frac{\ell_B}{\ell'} - \frac{x^2}{2} \right]}
L_{N-M}^{(M)}(x^2) \, x^M dx \:,
\end{multline}
where $L_{N-M}^{(M)}$ are the Laguerre polynomials. Note that \eqref{eq:PNM} (see also \eqref{eq:PNMa})
slightly differs from definition of similar integrals used in 
Refs.~\onlinecite{Kawabata,Kachorovskii,OurSSC} to simplify analysis of its properties
and numerical calculations, see Appendix~\ref{SEC:P}.

Then it is easy to show that by substitution 
\begin{equation}\label{eq:C_in_basis}
C_{\alpha}( {\bf{r}},{\bf{r}}' ) = 
W \sum_{N,k_y} \Phi_{N,k_y}({\bf{r}}) C_{\alpha N} \Phi_{N,k_y}^{\dag}({\bf{r}}') \:,
\end{equation}
we transform integral equations \eqref{eq:Cr012} into system of linear equations which allows 
to find Cooperons:
\begin{subequations}\label{eq:CN012}
\begin{equation}\label{eq:CN0}
   C_{0N} = 1 + \frac{\tau'}{\tau} P_N C_{0N},
\end{equation}
\begin{equation}\label{eq:CN1}
   C_{1N} = 1 + \frac{\tau'}{\tau} P_N C_{1N} + \frac{\tau'}{\tau_v} \Pi P_N C_{2N}
\end{equation}
\begin{equation}\label{eq:CN2}
   C_{2N} =  \frac{\tau}{\tau_v} \Pi + \frac{\tau'}{\tau} P_N C_{2N} + \frac{\tau'}{\tau_v} \Pi P_N C_{1N}
\end{equation}
\end{subequations}
where matrix $\Pi$ is a matrix $\Pi_v$ in the basis \eqref{eq:basisN}:
$ \Pi = \mathrm{diag} \left\{ 1, -1, 1, -1 \right\}$ with solution 
\begin{subequations}\label{eq:CN012res}
\begin{equation}\label{eq:CN0res}
   C_{0N} =  \left(1 - \frac{\tau'}{\tau} P_N\right)^{-1} ,
\end{equation}
\begin{equation}\label{eq:CN12res}
\begin{pmatrix} C_{1N} \\ C_{2N} \end{pmatrix} = 
\begin{pmatrix} 1-\frac{\tau'}{\tau}P_N & -\frac{\tau'}{\tau_v}\Pi P_N \\ 
    -\frac{\tau'}{\tau_v}\Pi & 1-\frac{\tau'}{\tau}P_N\end{pmatrix}^{-1}
\begin{pmatrix} 1 \\ \frac{\tau}{\tau_v} \Pi \end{pmatrix}
\end{equation}
\end{subequations}
Note that low magnetic field limit (see below) of these equations is exactly 
equal to non-diffusion approximation in zero magnetic field given in 
Ref.~\onlinecite{OurEPL}. 
Basis choice in Table~\ref{tbl:prod_basis} makes the form of all equations 
consistent with non-diffusion theory in zero magnetic field.\cite{OurEPL}
 
In the following, for the conductivity calculations we will need 
equations associated with fan diagrams starting 
from two or three scatterers, while our definition (see Fig.~\ref{fig:Ceq}), which is 
more convenient for Cooperon equation solution, gives sum of fan diagrams starting from
single scattering. To add one or two scatterings, one needs to multiply the result 
\eqref{eq:CN0res} to kernel of Cooperon equation:
\begin{subequations}\label{eq:CN012i}
\begin{equation}\label{eq:CN0i}
    C_{0N}^{(m)} =  \left(\frac{\tau'}{\tau} P_N\right)^{m-1} C_{0N},
\end{equation}
\begin{equation}\label{eq:CN12i}
    \begin{pmatrix} C_{1N}^{(m)} \\ C_{2N}^{(m)} \end{pmatrix} = 
\begin{pmatrix} \frac{\tau'}{\tau}P_N & \frac{\tau'}{\tau_v}\Pi P_N \\ 
    \frac{\tau'}{\tau_v}\Pi & \frac{\tau'}{\tau}P_N\end{pmatrix}^{m-1}
\begin{pmatrix} C_{1N} \\ C_{2N} \end{pmatrix}
\end{equation}
\end{subequations}
Equations \eqref{eq:CN012res},\eqref{eq:CN012i} allow easily compute 
Cooperons in the basis \eqref{eq:basisN}. 

\subsection{Conductivity correction}\label{SSEC:sigma}

\begin{figure}[t]
\hbox to \linewidth{
      \hfil
      \rlap{a)}\vtop{\vskip0pt\hbox{\includegraphics[width=0.45\linewidth]{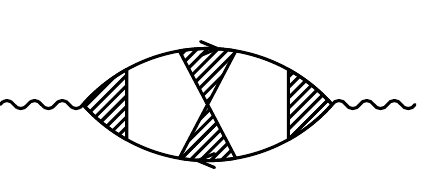}}}
      \hfil
      \rlap{b)}\vtop{\vskip0pt\hbox{\includegraphics[width=0.45\linewidth]{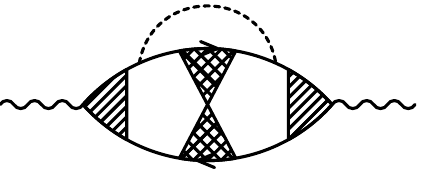}}}
      \hfil
}
\caption{Standard diagrams which give main contribution to the weak localization. 
}\label{fig:dia}
\end{figure}

A consistent theory of weak localization is
developed in the framework of the diagram technique. The
weak-localization corrections to the conductivity arise in
the first order of the parameter $(k_F\ell)^{-1}$. 
The weak localization correction to conductivity has two
contributions corresponding to standard diagrams illustrated in
Fig.~\ref{fig:dia} (see Refs.~\onlinecite{Zyuzin,Kachorovskii} for details).
It may be shown that all other diagrams either have higher in
$(k_F\ell)^{-1}$ order or do not depend on $\tau_{\phi}$ which defines the
magneto and temperature dependence of the conductivity
correction.

It may be shown that first diagram may be written as
\begin{equation}\label{eq:sigmaI_r}
\sigma_a = \frac{\hbar}{2\pi} \int  \operatorname{Tr} \left[ F ({\bvec{r}},{\bvec{r}}') \,
\mathcal{C}^{(3)}({\bf{r}}',{\bf{r}})  \right]
   d{\bf{r}}d{\bf{r}}' \:,
\end{equation}
where $F ({\bvec{r}},{\bvec{r}}') = J_x({\bvec{r}},{\bvec{r}}') \otimes J_x({\bvec{r}},{\bvec{r}}')$, 
$\bvec{J} ({\bvec{r}},{\bvec{r}}')$ is the combination of dressed current vertex with Green functions
$\bvec{J} ({\bvec{r}},{\bvec{r}}') = \int G(\bvec{r},\bvec{r}_1) \bvec{j}(\bvec{r}_1,\bvec{r}_2) G(\bvec{r}_2,\bvec{r}')$.
Bare vertex is velocity operator which may be written as 
\begin{equation}
    \bvec{j}_0 = e v \begin{pmatrix}
        \bgreek{\sigma} & 0 \\ 0 & -\bgreek{\sigma}
    \end{pmatrix} \delta(\bvec{r}-\bvec{r}')
\end{equation}
Evaluation analogous to one given in Ref.~\onlinecite{OurEPL} gives similar result
\begin{equation}\label{eq:vertex}
    \bvec{j} = 2 e v \begin{pmatrix}
        \bgreek{\sigma} & 0 \\ 0 & -\bgreek{\sigma}
    \end{pmatrix} \delta(\bvec{r}-\bvec{r}')
\end{equation}
It should be noted that \eqref{eq:vertex} assumes $\tau_{\phi} \gg \tau$, 
but has no restrictions on $\tau_v$ compared with $\tau$.

Direct evaluation with the help of equations given in  Appendix \ref{SEC:int}
for a dressed current leads to
\begin{equation}
    J^{\pm} ({\bvec{r}},{\bvec{r}}') = \pm e\frac{\sqrt2 \ell'}{\hbar} n_{\pm}
    \left[ G^R({\bvec{r}},{\bvec{r}}') +  G^A({\bvec{r}},{\bvec{r}}') \right]
\end{equation}

To write conductivity correction using solution of Cooperon equation obtained earlier
we should rewrite $F(\bvec{r},\bvec{r}')$ in the ``product space basis''. 
In the following it is important to note that $F$ in the conductivity correction 
is integrated and in the product terms 
$\left[G^{R,A}({\bvec{r}},{\bvec{r}}')\right]^2$ are small compared to 
$G^{R/A}({\bvec{r}},{\bvec{r}}')G^{A/R}({\bvec{r}},{\bvec{r}}')$ because of the smallness 
of $(k_F\ell)^{-1}$.

Also it is important that current has a $n_{\pm}$ as a multiplicator. It leads to 
zero off-diagonal components of current $\sigma_{+-}=\sigma_{-+}=0$ 

Finally, we may rewrite product of two Green functions using \eqref{eq:P_def} which gives
\begin{multline}
    \sigma_a = - \frac{e^2 {\ell'}^2}{\pi\hbar} \sum_{\alpha\beta\gamma\delta} 
    \int \frac1{2W}\Big[ 
    P^{\delta\alpha}_{\beta\gamma}({\bvec{r}},{\bvec{r}}')  \\
    \left.+P _{\delta\alpha}^{\beta\gamma}({\bvec{r}},{\bvec{r}}')
    \right] \left\{ \mathcal{C}^{(3)} \right\}^{\alpha\beta}_{\gamma\delta}({\bf{r}}',{\bf{r}}) 
\end{multline}
which with the definition
\begin{equation}
Q^{\beta\alpha}_{\delta\gamma} = -
\frac{P^{\delta\alpha}_{\beta\gamma}({\bvec{r}},{\bvec{r}}') + P_{\delta\alpha}^{\beta\gamma}({\bvec{r}},{\bvec{r}}')}2
\end{equation}
may be written as 
\begin{equation}
    \sigma_a = \frac{e^2 {\ell'}^2}{\pi\hbar} \sum_{\alpha\beta\gamma\delta} 
    \int \frac1{W} Q^{\beta\alpha}_{\delta\gamma}({\bvec{r}},{\bvec{r}}') 
    \left\{ \mathcal{C}^{(3)} \right\}^{\alpha\beta}_{\gamma\delta}({\bf{r}}',{\bf{r}}).
\end{equation}
Exchange of couple left(right) indices in one-particle space in the product space 
chose in accordance with the Table~\ref{tbl:prod_basis} is 
equivalent to multiplication from the left (right) to the matrix
\begin{equation}
    \Pi_* = \begin{pmatrix}
        \Pi_r & 0 & 0 & 0 \\
        0 & 0 & \Pi_r & 0 \\
        0 & \Pi_r & 0 & 0 \\
        0 & 0 & 0 & \Pi_r
    \end{pmatrix},
\end{equation}
where 
\begin{equation}
    \Pi_r = \begin{pmatrix}
        1 & 0 & 0 & 0 \\
        0 & 0 & 1 & 0 \\
        0 & 1 & 0 & 0 \\
        0 & 0 & 0 & 1
    \end{pmatrix}.
\end{equation}
It is then essential to take advantage from the block structure of 
$P({\bvec{r}},{\bvec{r}}')$ and $\Pi_*$ which allows one to 
rewrite $Q$ in a product space as 
\begin{equation}
    Q({\bf{r}},{\bf{r}}') = \begin{pmatrix}
        Q_{sv}({\bf{r}},{\bf{r}}') & 0 & 0 & 0 \\
        0 & 0 & Q_{sv}({\bf{r}},{\bf{r}}') & 0 \\
        0 & Q_{sv}({\bf{r}},{\bf{r}}') & 0 & 0 \\
        0 & 0 & 0 & Q_{sv}({\bf{r}},{\bf{r}}')
    \end{pmatrix},
\end{equation}
with
\begin{equation}
    Q_{sv}({\bf{r}},{\bf{r}}') = - \frac{\Pi_r P_{sv}({\bf{r}},{\bf{r}}') + P_{sv}({\bf{r}},{\bf{r}}')\Pi_r }2.
\end{equation}
Then we use block form if the Cooperons \eqref{eq:Cr} and rewrite these equations as 
\begin{equation}\label{eq:Sar}
    \sigma_a = 2\frac{e^2 }{\pi\hbar}\frac{{\ell'}^2}{W} 
    \int {\mathrm{Tr}} \left\{ Q_{sv}({\bf{r}},{\bf{r}}')
    \left[ C_0({\bf{r}}',{\bf{r}})  - C_2({\bf{r}}',{\bf{r}})  \right] \right\}
\end{equation}

For the computation, we decompose \eqref{eq:Sar} in the basis \eqref{eq:basisN}:
\begin{equation}\label{eq:Sar_basisN}
    Q_{sv}({\bf{r}},{\bf{r}}') = 
    \sum_{N,k} \Phi_{Nk}({\bf{r}}) Q_N \Phi_{Nk}^{\dag}({\bf{r}}') \:,
\end{equation}
It may be shown that 
\begin{equation}
    Q_N = \frac{\Pi' P_N + P_N \Pi'}2
\end{equation}
Where $\Pi'$ is the matrix similar to $\Pi$ defined in Eq.~\eqref{eq:CN012}, 
it is a $-\Pi_r$ in the basis \eqref{eq:basisN}: 
$ \Pi' = \mathrm{diag} \left\{ -1, -1, 1, -1 \right\} $.
Note opposite sign which we introduced to make $\Pi'$ and $\Pi$ 
conincide with matrix defined in Ref.~\onlinecite{OurEPL}.

And  the final result for the weak localization correction associated with 
diagram Fig.~\ref{fig:dia}(a) is given by
\begin{equation}\label{eq:Sa}
    \sigma_a = 2\frac{e^2 }{\pi^2\hbar}\frac{2{\ell'}^2}{\ell_B^2} 
    \sum_N {\mathrm{Tr}} \left\{ Q_N \left[ C_{0N}^{(3)}  - C_{2N}^{(3)} \right] \right\}.
\end{equation}

Note that the result \eqref{eq:Sa} is extremely similar to weak localization correction 
in non-diffusion theory in zero magnetic field. Moreover, accurate calculation of zero 
magnetic field limit as a formal limit $\ell/\ell_B \to 0$ gives the result which is exactly 
equal to results presented in Ref.~\onlinecite{OurEPL}.
The notation in the current manuscript is chosen to simplify this comparison.
As explained in Appendix~\ref{SEC:sum}, in addition it facilitates the use of
low field limit for precise computation of slowly converging infinite sum \eqref{eq:Sa}.

For the diagrams of type (b) evaluation of conductivity correction is more complicated. 
We will formulate it using definition
\begin{equation}
    L^{\pm}(\bvec{r},\bvec{r}') = \pm n_{\pm} P(\bvec{r},\bvec{r}')
\end{equation}
which originates from combination of dressed current and Green function
\begin{multline}
    e\frac{\sqrt2 \ell'}{\hbar} \left\{ L ^j \right\}^{\alpha\beta}_{\gamma\delta}(\bvec{r},\bvec{r}') \simeq 
    J_{\alpha\beta}^j(\bvec{r},\bvec{r}') G^R_{\gamma\delta}(\bvec{r},\bvec{r}')  \\ \simeq 
    J_{\gamma\delta}^j(\bvec{r},\bvec{r}') G^A_{\alpha\beta}(\bvec{r},\bvec{r}') 
\end{multline}
where sign $\simeq$ is used to show the equivalence up to terms proportional 
to $(k_F\ell)^{-1}$ in the conductivity.

With this definition conductivity may be written as (note that there are 
two diagrams with scattering "above" and "below" Cooperon) 

\begin{widetext}
\begin{equation}
    \sigma_b^{ij} = \frac{e^2}{2\pi} \frac{2{\ell'}^2}{\hbar^2} 
    \int {\mathrm{Tr}} \left\{ L^{i}(\bvec{r}',\bvec{r}_1) S \Pi_* 
    L^{j}(\bvec{r}_1,\bvec{r}) C^{(2)}(\bvec{r},\bvec{r}')
    +L^{j}(\bvec{r}',\bvec{r}_1) S \Pi_* 
    L^{i}(\bvec{r}_1,\bvec{r}) C^{(2)}(\bvec{r},\bvec{r}')
    \right\}
\end{equation}

By using block form of matrices we may write this result as 
\begin{multline}
    \sigma_b^{ij} = 2 \frac{e^2}{\pi} \frac{{\ell'}^2}{\hbar^2} 
    \int {\mathrm{Tr}} \bigg\{ 
    L^{i}_{sv}(\bvec{r}',\bvec{r}_1) \Pi_r L^{j}_{sv}(\bvec{r}_1,\bvec{r}) 
    \left[C_0(\bvec{r},\bvec{r}') - C_2(\bvec{r},\bvec{r}') \right]
  \\ 
    - \frac{\tau}{\tau_v}  L^{i}_{sv}(\bvec{r}',\bvec{r}_1) 
    \Pi_v \Pi_r L^{j}_{sv}(\bvec{r}_1,\bvec{r}) C_1(\bvec{r},\bvec{r}')
    \bigg\} + i\leftrightarrow j
\end{multline}
\end{widetext}

For the computation, we rewrite $L^{\pm}$ in the basis \eqref{eq:basisN}. 
Direct calculation with the use of Appendix \ref{SEC:int} gives
\begin{subequations}\label{eq:Ldecomp}
\begin{multline}
    L^+_{sv} (\bvec{r},\bvec{r}') = i \sum_k \Bigg[ 
   \Psi_{0k}(\bvec{r}) L_0^T \Psi_{0k}^{\dag}(\bvec{r}') + \\
   \sum_N  \Psi_{N+1,k}(\bvec{r}) L_{N+1}^T \Psi_{Nk}^{\dag}(\bvec{r}') 
   \Bigg]
\end{multline}
\begin{multline}
    L^-_{sv} (\bvec{r},\bvec{r}') = i \sum_k \Bigg[ 
   \Psi_{0k}(\bvec{r}) L_0 \Psi_{0k}^{\dag}(\bvec{r}') + \\
   \sum_N  \Psi_{Nk}(\bvec{r}) L_{N+1} \Psi_{N+1,k}^{\dag}(\bvec{r}') 
   \Bigg]
\end{multline}
\end{subequations}
where we defined (technically, we have to compute $L_1$ separately, but 
the result is the same as for $L_{N\ge2}$ setting $P^M_N=0$
for $M>N$)
\begin{subequations}\label{eq:LN}
\begin{equation}
    L_{N\ge1} = \epsilon \begin{pmatrix}
        0 & 0  & 0 & 0 \\
        0 & -iP_{N-1}^1 & -P_{N}^2 & iP_{N+1}^3 \\
        0 & P_{N-1}^0 & -iP_N^1 & -P_{N+1}^2 \\
        0 & -iP_{N}^1 & P_{N}^0 & -iP_{N+1}^1 
    \end{pmatrix} \epsilon
\end{equation}
\begin{equation}
    L_{0} =  \epsilon \begin{pmatrix}
        0 & 0 & 0 & 0 \\
        0 & 0 & -P_{0}^0 & iP_{1}^1 \\
        0 & 0 & 0 & 0 \\
        0 & 0 & 0 & 0 
    \end{pmatrix} \epsilon
\end{equation}
\end{subequations}

One may note that $\sigma_b^{++}=\sigma_b^{--}=0$ and 
$\sigma_b^{+-}=\sigma_b^{-+}\equiv \sigma_b$.

Analogously to $\sigma_a$, we arrive to the following equation
for $\sigma_b$:
\begin{multline}\label{eq:Sb}
    \sigma_b = \frac{e^2 }{\pi^2\hbar}\frac{2{\ell'}^2}{\ell_B^2} 
    {\mathrm{Tr}} \bigg\{ \\
    L_0\Pi'L_0^T \left[ C_{00}^{(2)}  - C_{20}^{(2)} \right]
    -\frac{\tau}{\tau_v} L_0 \Pi\Pi' L_0^T C_{10}^{(2)}
    \\
    + \sum_N 
    \left[ L_N^T \Pi' L_N +L_{N+1} \Pi' L_{N+1}^T \right]
    \left[ C_{0N}^{(2)}  - C_{2N}^{(2)} \right] 
    \\
    - \frac{\tau}{\tau_v} \left[L_N^T \Pi\Pi' L_N + L_{N+1} \Pi\Pi' L_{N+1}^T\right]C_{1N}^{(2)}
    \bigg\}
\end{multline}

The structure of the answer is very close to the one obtained in Ref.~\onlinecite{OurEPL} with
integration is replaced with a summation over Landau levels. The only significant 
discrepancy is an extra term for 0-th Landau level. 

Note that for a chosen scattering contribution to the conductivity from the singlet is zero. 
Practical computation may assume all computations for $3\times 3$ block in matrices. 
For completeness we left singlet contribution in all equations to avoid possible confusion.

\subsection{Low magnetic field limit}\label{SSEC:LowH}
In the above, we assumed classically weak magnetic fields $k_F\ell_B\gg1$. 
This allowed us following Kawabata\cite{Kawabata} to factorize electron Green function into 
zero-field Green function and phase factor which absorbs the effect of magnetic field.

Diffusion pole in Cooperon equation is cut off by phase relaxation time which means that
infinite sum in \eqref{eq:Sa}, \eqref{eq:Sb} is defined by $N\gg1$ and $N$ may be 
correctly replaced with integration over continuous variable if 
$\ell^2/2\ell_B^2\gg \tau_{\phi}/\tau$
Only for such small magnetic fields the we may replace summation with the 
integration assuming $N$ large. 

In this case, we may find $P_N^M$ \eqref{eq:PNM} in elementary functions 
by using Mehler-Heine asymptotic for Laguerre polynomials
\begin{equation}\label{eq:LJ}
    L_{N}^{(M)}(x\to 0)\rightarrow e^{\frac{x}2}\left(\frac{N}{x} \right)^{\frac{M}2}J_{M}\left( 2\sqrt{xN}\right)
\end{equation}
\begin{equation}
\begin{split}
    P_N^M(\alpha) = (-i)^M I_M\left( \frac{2\sqrt{N}}{\alpha} \right),\\
    I_M(q) = \frac{\left[\sqrt{1+q^2}-1\right]^M}{q^M\sqrt{1+q^2}}
\end{split}
\end{equation}

And the summation may be replaced with integration using the following rule:
\begin{equation}
    \sum_N f(N) \longrightarrow    
    \frac{\ell_B^2}{2}\int_0^{\infty} f\left( N=q^2\ell_B^2/4 \right) qdq 
\end{equation}

Following this procedure, we may obtain non-diffusion results for the weak localization
correction \cite{OurEPL} from \eqref{eq:Sa}, \eqref{eq:Sb}.

\section{Results}\label{SEC:res}
\begin{figure}[t]
\includegraphics[width=\linewidth]{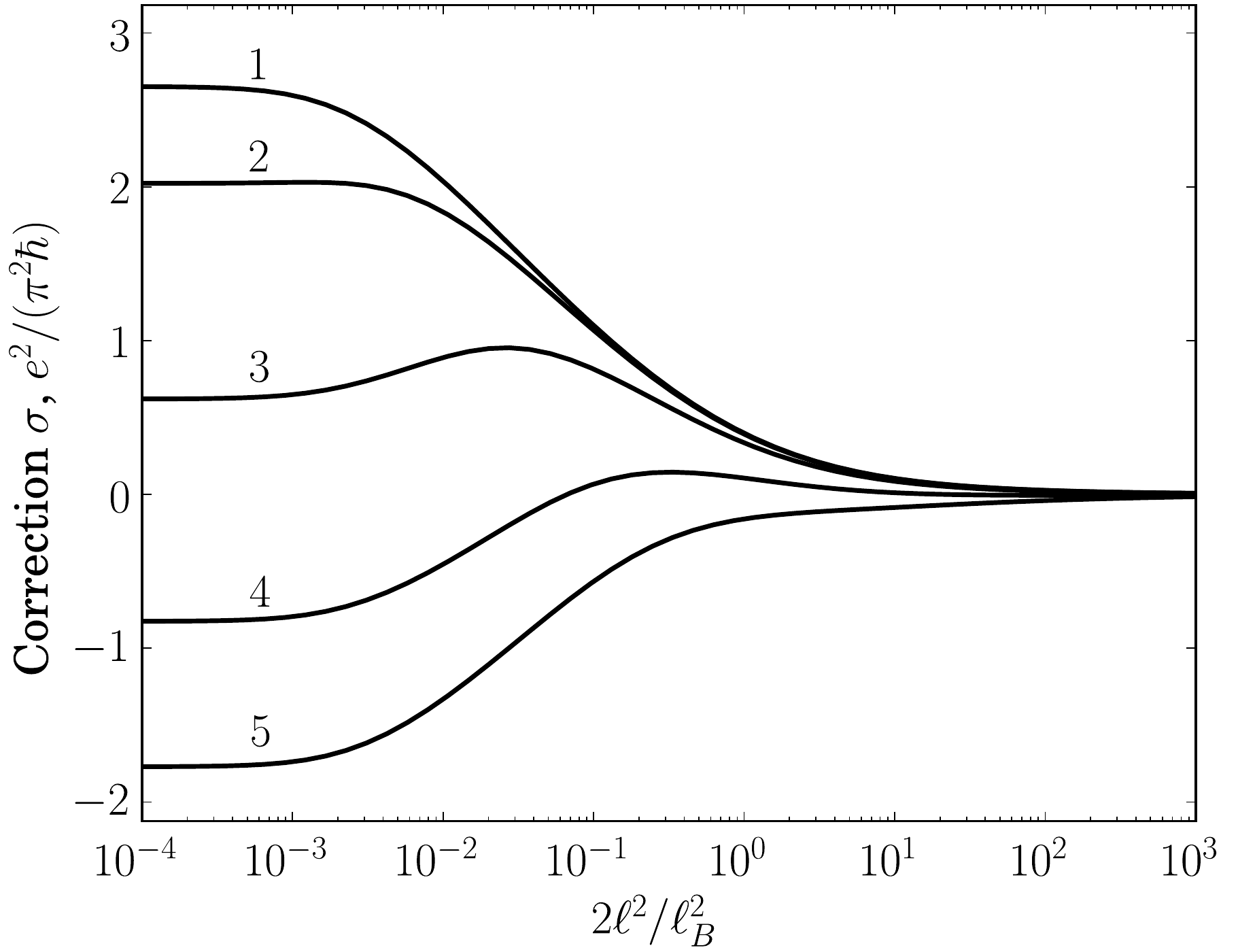}
\caption{Weak localization correction as a function of magnetic field.
Here $\frac{\tau}{\tau_{\phi}}=10^{-3}$ and 
$\frac{\tau}{\tau_v}=0.0,\;10^{-3},\;10^{-2},\;0.05,\;0.1$ 
(labelled as 1, 2, 3, 4, 5 correspondingly).
}\label{fig:WL}
\end{figure}

Fig.~\ref{fig:WL} is the main result of our work and it shows the transition from
weak antilocalization to the weak localizations when changing the intervalley 
transitions rate. Curve 1 corresponds to the absence of intervalley transitions,
in this case zero field contribution to the quantum correction is positive (WAL) 
and conduction decreases as a function of magnetic field. 
Curve 2 corresponds to the case when intervalley transitions time is 
equal to the phase decoherence time in one Dirac cone. In this case 
WAL survives, but quantum correction decreases. Increase of 
intervalley transitions as compared to loss of coherency in one valley
leads to further suppression of the quantum correction in zero magnetic 
field and to non monotonous dependence of the conductivity as a function 
of magnetic field. The latter is caused by the fact that when mahnetic 
length $\ell_B$ becomes comparable with diffusion length $\ell$, 
the WAL correction is suppressed in comparison with the decoherence time 
in one valley. Further decrease of intervalley transitions in 50 and 100
times as compared to $\tau_{\phi}$ leads to the change of quantum 
correction sign which results in ``restore'' of the time-reversal symmetry 
and necessity to take into account states of the both valleys when 
considering the carriers scattering. 
Curve 4 has an extremum point yet, but curve 5 finally corresponds to 
``conventinoal'' magnetoresistance caused by the weak localization.

Intervalley contribution to the conductivity due to its time-invariant nature
results in a conventional WL, similar to spinless single valley case. 
However, it is proportional to the intervalley scattering rate.
Without intervalley scattering, when trigonal warping and reduced symmetry 
scatterers may be neglected, the intravalley contribution comes into play. 
Due to Berry phase which originates from the structure of the wavefunctions
intravalley scattering introduces additional phase which results in opposite 
sign of conductivity correction.\cite{Ando02,OurSSC,Kechedzhi07}
Controlling the intervalley scattering rate one may continuously switch between
two cases. 

Results presented in Fig.~\ref{fig:WL} demonstrate that the change of 
localization correction sign takes place at relatively effective intervalley
transitions, when the phase relaxation time in one valley 
is about two orders of magnitude longer than these transitions, 
but the condition $\tau_{\phi}>\tau_v \gg \tau$ still holds.
In the Ref.~\onlinecite{Tikhonenko07} the transition from WL to WAL 
has been observed in graphene when the carrier concentration has been 
decreased. In the framework of the theory developed here these results
may be interpreted as a change of the ratio $\tau_{\phi}/\tau_v$
as a function of carrier concentration.

Non monotonous dependence of conductivity as a function of magnetic 
field (curve 3) for a region of $\tau_v/\tau_{\phi}$ ratio is a 
characteristic property of two-dimensional graphene layers defined by
its band structure. Similar behaviour of $\sigma(B)$ has not been 
observed in other multivalley systems, e.g. Si.\cite{Tarasenko07}

To conclude, we have developed the theory of the weak localization in 
graphene with account on intervalley transitions in the whole range of
classically weak
magnetic fields. 
It has been shown that in two-dimensional graphene layers
intervalley transitions lead to crossover between weak antilocalization 
and  weak localization regimes and all peculiarities of the system are 
defined by spinor-like character of carrier wavefunctions. 
Spin-orbit interaction and suppression of backscattering in quantum
relaxation time are shown to be virtually unimportant 
in the theoretical description of the weak localization in graphene.
In addition, we have developed a novel analytical approach for the 
calculation of weak localization in the systems with linear 
spectrum which allows to simplify conceptually the numerical 
calculation of quantum correction to the conductivity in 
non-zero magnetic fields.

\appendix
\section{Numerical calculation of $P_N^M(x)$}\label{SEC:P}
In this section we comment on numerical computation of integrals (in this section we 
sometimes omit function argument $\alpha$ for brevity)
\begin{equation}\label{eq:PNMa}
 P_N^M(\alpha)=\alpha\sqrt{\frac{(N-M)!}{N!}}\int e^{-\alpha \zeta -\frac12 \zeta^2}
   \zeta^M L_{N-M}^M \left( \zeta^2 \right) d\zeta.
\end{equation}
Laguerre polynomials for large $N$ are highly oscillating functions, and straightforward 
numerical integration using \eqref{eq:PNMa} is impractical.
Moreover, if one wants to use this definition to compute $P_N^M(\alpha)$, for large 
integration variable exponent underflows and Laguerre polynomial overflows, despite the fact that 
their product is finite and well defined. The latter problem may be overcame by redefining integrand 
via Laguerre function $e^{-x/2}x^{M/2}L_{N-M}^{M}(x)$ and calculating this function using 
recurrent relation which may be easily derived from recurrence relations for Laguerre polynomials.
However, there exists a more efficient approach which allows to obtain the functions 
\eqref{eq:PNMa} with negligible computational complexity.

As shown in Kawabata's work,\cite{Kawabata} Laguerre integrals \eqref{eq:PNMa}
with $M=0$ satisfy four-term recurrence relation
\begin{subequations}\label{eq:P0rec}
\begin{align}
P_0^0(\alpha)&=\alpha \sqrt{\frac{\pi}2} 
  \operatorname{exp}\left( \frac{\alpha^2}2 \right) \operatorname{erfc} \left( \frac{\alpha}{\sqrt2}\right),\\
P_1^0(\alpha)&=\alpha^2 - \alpha^2 P_0(\alpha), \\
P_2^0(\alpha)&=\frac{1+\alpha^2}2 \left[ P_0(\alpha) - P_1(\alpha) \right], \\
P_N^0(\alpha)&=\frac{N-2}{N}P_{N-3}(\alpha) \nonumber \\
 & + \frac{N-1+\alpha^2}{N} \left[ P_{N-2}(\alpha) - P_{N-1}(\alpha) \right].
\end{align}
\end{subequations}
It may be shown that this recurrence relation is numerically unstable.
For computations one may rewrite recurrence relation \eqref{eq:P0rec} as a 
system of linear equations with banded matrix\footnote{Note the use of 
$P_1^0$ and $P_2^0$ as a left boundary condition. Surprisingly, this strongly 
improves precision of the result. Also, in the functions $P_1^0(\alpha)$ and 
$P_2^0(\alpha)$ at the left boundary 
for large argument instead of using \eqref{eq:P0rec} one should use series 
in $1/\alpha$ for a reasonable accuracy.}
\begin{widetext}
\begin{equation}\label{eq:P0rec_mat}
    \begin{pmatrix}
        a_4  &  1  &    &  &  &  \\
       -a_5  & a_5 &  1  &  &  &  \\
        b_6  &-a_6 & a_6 & 1 &   &  \\
             & \ddots    & \ddots &\ddots & \ddots & \\
             &    &  b_{N_{max}-1} & -a_{N_{max}-1} & a_{N_{max}-1} & 1 \\
             &    &   & b_{N_{max}} & -a_{N_{max}} & a_{N_{max}} \\
    \end{pmatrix}
    \begin{pmatrix} 
        P_3^0 \\ P_4^0 \\ P_5^0 \\ \vdots \\ P_{N_{max}-2}^0 \\ P_{N_{max}-1}^0 
    \end{pmatrix}
    =
    \begin{pmatrix} 
        - b_4 P_1^0 + a_4 P_2^0 \\ -b_5 P_2^0 \\ 0 \\ \vdots \\ 0\\ -P_{N_{max}}^0 
    \end{pmatrix},
\end{equation}
\end{widetext}
where
\begin{equation}
    a_n = \frac{n-1+\alpha^2}{n},\;\;\; b_n = -\frac{n-2}{n},
\end{equation}
and $P_{N_{max}}^0$ may be taken from approximation for large $N$
\begin{equation}\label{P_lim}
    P_N^0(\alpha) \stackrel{N\to\infty}{\longrightarrow} \frac{1}{\sqrt{1+\frac{4(N+1/2)}{\alpha^2}}}.
\end{equation}
Note that the numerical error due to use of approximated value for $P_{N_{max}}^0$ exponentially 
decays for small $N$ and the result is exact for $N<N_{max}-\delta N$ where 
$\delta N$ is relatively small.

For $M\neq 0$ we define scaled integrals 
\begin{subequations}\label{eq:P123def}
\begin{align}
    {P}_n^1 & = \frac{\alpha}{\sqrt{n}}\tilde{P}_{n}^1, \\
    {P}_n^2 & = \frac{1     }{\sqrt{(n-1)n}}\tilde{P}_{n}^2, \\
    {P}_n^3 & = \frac{\alpha}{\sqrt{(n-2)(n-1)n}}\tilde{P}_{n}^3,
\end{align}
\end{subequations}
which satisfy the following recurrences:
\begin{subequations}\label{eq:P123rec}
\begin{align}
    \tilde{P}_n^1 & = \tilde{P}_{n-2}^1 - P_{n-1}^0 + P_{n-2}^0,\\
    \tilde{P}_n^2 & = 1 + \tilde{P}_{n-2}^2 - 
    (1+\alpha^2) \tilde{P}_{n-1}^1 - (1-\alpha^2) \tilde{P}_{n-2}^1, \\
    \tilde{P}_n^3 & = \tilde{P}_{n-2}^3  + 2\tilde{P}_{n-2}^1 - 
    \tilde{P}_{n-1}^2 + \tilde{P}_{n-2}^2.
\end{align}
with additional initial values
\begin{align}
    \tilde{P}_1^1 & = - P_1^0 + P_0^0,\\
    \tilde{P}_{n<M}^{M} & = 0.
\end{align}
\end{subequations}

It is worth to note that approach (\ref{eq:P0rec_mat}-\ref{eq:P123rec}) needs only $\mathcal{O}(N)$ floating point 
operations to compute $4N$ functions $P_{n<N}^{M=0,1,2,3}(\alpha)$ for each argument $\alpha$.
This is negligible compared with $\mathcal{O}\left(N^3\right)$ operations to compute them using \eqref{eq:PNMa}.
If one is interested in benchmarking recurrent approach (\ref{eq:P0rec_mat}-\ref{eq:P123rec}), it 
is practical to compare it with ``reference'' values calculated by 
definition \eqref{eq:PNMa} (taking into account remark about 
the Laguerre function computation) to estimate necessary $N_{max}$ and 
$\delta N$ as a functions of $\alpha$ to obtain any desired accuracy. 

\section{Numerical calculation of infinite sums in conductivity}\label{SEC:sum}

\begin{figure}[t]
\includegraphics[width=\linewidth]{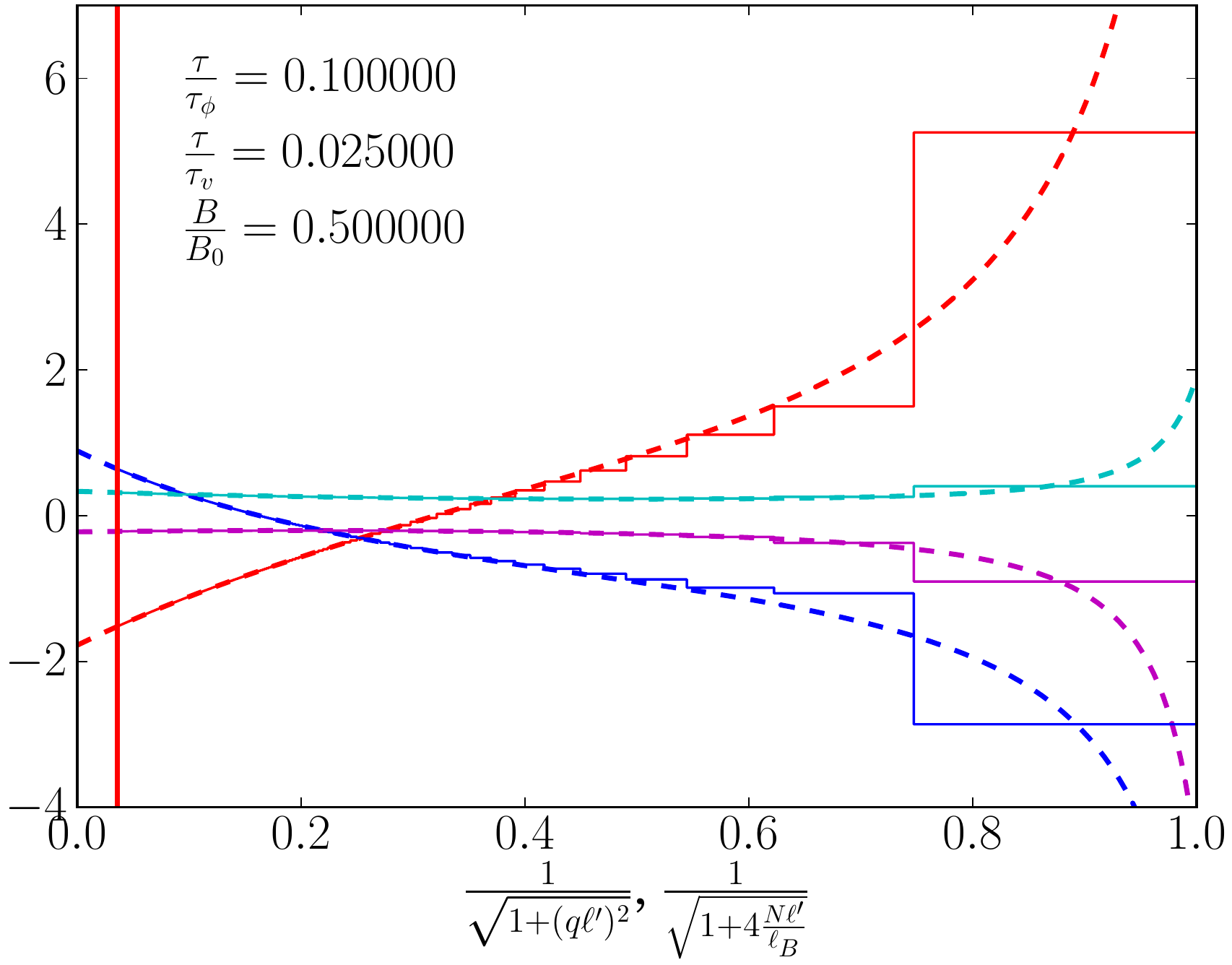}
\caption{Illustration of accurate summation by using 
nondiffusion approximation for large values of Landau level number.
Contributions to the weak localization correction calculated using 
\eqref{eq:Sa}, \eqref{eq:Sb} are 
compared with non-diffusion approximation \cite{OurEPL} after changing 
integration variable to $I=1/\sqrt{1+(q\ell')^2}$.
For comparison, we treat sum as an integral of almost everywhere constant 
function of real variable $N$ and change the variable to 
$1/\sqrt{1+(2N\ell'/\ell_B)^2}$.
Intravalley and intervalley contributions from diagrams of type (a) and (b) are 
shown separately.
Vertical line shows position of 1000th Landau level. For an accurate computation, we 
use sum for $N<1000$ and integral for $N\ge 1000$
}\label{fig:N_int}
\end{figure}

Conductivity correction written as a sum over Landau level converges extremely slowly, it is easy 
to show that 
\begin{equation}
    \sum_{N_{max}}^{\infty} \sim \mathcal{O}\left( \frac1{\sqrt{N_{max}}}\right),
\end{equation}
which makes it technically complicated to evaluate sums in Eqs.~\eqref{eq:Sa}, \eqref{eq:Sb}
with reasonable precision. 
For realistic calculations the convergence of these sums may be significantly improved: 
if the infinite sum remainder is approximated by the low magnetic field limit, 
then it is easy to show that 
\begin{equation}
    \sum_{N_{max}}^{\infty} - \int_{N_{max}}^{\infty} \sim \mathcal{O}\left( \frac1{N_{max}}\right).
\end{equation}
This crucially improves convergence of infinite sums.
Similar approach is mentioned briefly in Ref.~\onlinecite{Cassam}. 

To illustrate this approximation, in Fig.~\ref{fig:N_int} we show weak localization correction
contributions separately for $\tau/{\tau_{\phi}}=0.1$, $\tau/\tau_v=0.025$, $B/B_0=2\ell^2/\ell_B^2=0.5$. 
To simplify the comparison the sums \eqref{eq:Sa}, \eqref{eq:Sb} may be written as integrals of
piecewise constant functions of continuous $N$. These functions are shown in Fig.~\ref{fig:N_int}
in solid lines and their approximation with low magnetic field limit (see Sec.~\ref{SSEC:LowH}) 
are shown in dashed lines. 

Mathematically, it is clear that the low field limit is based not on the 
large compared with unity value of $N\ell/\ell_B$. For any value of magnetic field, there 
exists a number $N_{max}$ when \eqref{eq:LJ} gives a good approximation of Laguerre polynomial and 
summation may be replaced with the integration.

Physically, effect of magnetic field on large trajectories is always diffusion-like. It is a 
matter of the size of a trajectory compared with magnetic length, when the trajectory may be 
considered ``large''.

\section{Some integral and sums}\label{SEC:int}
In the manuscript, we used some mathematical facts which are given here for 
completeness.
In the manuscript we use definitions:
\begin{equation}
    \bgreek{\rho}=\bvec{r}-\bvec{r}' \;, \;\;\;
    n_{\pm} = \frac{\rho_x \pm i\rho_y}{|\bgreek{\rho}|}
\end{equation}
All infinite sums for single-particle basis functions may be derived from the following 
relation for oscillator functions \eqref{eq:osc_fun}:
\begin{multline}\label{eq:sum_N_osc}
    \sum_{N,k} \frac{\psi_{N,k}(\bvec{r})\psi^*_{N-\eta M,k}(\bvec{r}')}
    {\varepsilon - \hbar \omega_c \sqrt{N} \pm \frac{i\hbar}{2\tau'}} \simeq  \\ - 
    \frac{ \left( \mp i\eta n_{\eta} \right)^M e^{-i\frac{(x+x')(y-y')}{2\ell_B^2}}}
    {\hbar v \sqrt{2\pi} \sqrt{\rho/k_F}}  e^{-\frac{\rho}{2\ell'}} e^{\pm i\left( \rho k_F + \pi/4 \right)}.
\end{multline}
Here $\eta=\pm1$.

Equation \eqref{eq:sum_N_osc} may be obtained as a limit $k_F\ell \gg 1$, 
$k_F\ell_B \gg 1$ of the exact relation 
\begin{multline}
    \sum_{N,k} \frac{\psi_{N,k}(\bvec{r})\psi^*_{N-\eta M,k}(\bvec{r}')}
    {\varepsilon - \hbar \omega_c \sqrt{N} \pm \frac{i\hbar}{2\tau'}} = \\  
    - \frac{e^{-i\frac{(x+x')(y-y')}{2\ell_B^2}}e^{-\frac{\rho^2}{4\ell_B^2}}
    \left( \frac{\rho^2}{2\ell_B^2} \right)^{\frac{M}2}}{2\pi\hbar v} n_{\eta}^M 
    \times \\
    \frac1{\ell_B} \sum_N \sqrt{\frac{(N-M)!}{N!}}\frac{L_{N-M}^M\left( \frac{\rho^2}{2\ell_B^2} \right)}
    {2\sqrt{N}- \left( k_F\ell_B \pm i\frac{\ell_B}{2\ell'} \right)}
\end{multline}
under additional assumptions $M\sim 1$, $\rho \ll \ell_B^2 k_F$.

Matrix elements of Cooperon kernel in the basis two-particle functions may be derived from
matrix elements of $P_0\left(\bvec{r},\bvec{r}'\right)$ \eqref{eq:P0} which are obtained by 
direct calculation
\begin{multline}
    \int \phi_{Nk}^*(\bvec{r}) n_{\pm}^M P_0(\bvec{r},\bvec{r}') \phi_{N'k'}(\bvec{r}') 
    = \\
    \frac{\ell'}{\ell} \delta_{k,k'}\delta_{N,N'\pm M} (\pm1)^M
    P_{\mathrm{max}\left\{N,N'\right\}}^M \left( \frac{\ell_B}{\ell'} \right)
\end{multline}
where $P_N^M$ defined in \eqref{eq:PNMa}

\section*{Acknowledgments} 
The authors acknowledge fruitful discussions with L.E.~Golub and S.A.~Tarasenko. 
This work was supported by the RFBR
grant 12-02-00580, EU project ``POLAPHEN'',
RF President Grant NSh-1085.2014.2 and 
by the Goverment of Russia through the program 
P220 (project 14.Z50.31.0021, leading scientist M. Bayer).

\bibliography{inter_WL}

\end{document}